\documentclass[showpacs,amsmath,amssymb,amsfonts,twocolumn,superscriptaddress,checklength]{revtex4-1}
\usepackage{graphicx}
\usepackage{caption}
\usepackage{subfigure}
\usepackage{verbatim}
\usepackage{dcolumn}% Align table columns on decimal point
\usepackage{bm}% bold math
\usepackage{epsf}
\usepackage{xcolor}
\usepackage{hyperref}
\usepackage{txfonts}%
\usepackage{enumerate}
\usepackage{bbm}
\usepackage{lipsum}
\usepackage{amsmath}
\usepackage{amsfonts}
\hypersetup{
	colorlinks=true,   % Options: false (boxed links); true (colored links).
	linkcolor=blue,  % Color of internal links.
	citecolor=blue, % Color of links to bibliography.
	urlcolor =blue    % Color of external links.
}
\usepackage[justification=centering,singlelinecheck=false]{caption}
\begin{document}
\title{Unidirectional perfect absorption induced by chiral coupling in spin-momentum locked waveguide magnonics}

\author{Jie Qian}
\affiliation{Zhejiang Key Laboratory of Micro-Nano Quantum Chips and Quantum Control, School of Physics, State Key Laboratory for Extreme Photonics and Instrumentation, Zhejiang University, Hangzhou 310027, China}
\affiliation{State Key Laboratory of Precision Spectroscopy, Institute of Quantum Science and Precision Measurement,East China Normal University, Shanghai 200062, China}
\author{Qi Hong}
\affiliation{Zhejiang Key Laboratory of Micro-Nano Quantum Chips and Quantum Control, School of Physics, State Key Laboratory for Extreme Photonics and Instrumentation, Zhejiang University, Hangzhou 310027, China}
\author{Zi-Yuan Wang}
\affiliation{Zhejiang Key Laboratory of Micro-Nano Quantum Chips and Quantum Control, School of Physics, State Key Laboratory for Extreme Photonics and Instrumentation, Zhejiang University, Hangzhou 310027, China}
\author{Wen-Xin Wu}
\affiliation{Zhejiang Key Laboratory of Micro-Nano Quantum Chips and Quantum Control, School of Physics, State Key Laboratory for Extreme Photonics and Instrumentation, Zhejiang University, Hangzhou 310027, China}
\author{Yihao Yang}
\affiliation{State Key Laboratory of Extreme Photonics and Instrumentation, ZJU-Hangzhou Global Scientific and Technological Innovation Center, Zhejiang University, Hangzhou 310027, China}
\author{C.-M. Hu}
\email{canming.hu@umanitoba.ca}
\affiliation{Department of Physics and Astronomy, University of Manitoba, Winnipeg R3T 2N2, Canada}
\author{J. Q. You}
\email{jqyou@zju.edu.cn}
\affiliation{Zhejiang Key Laboratory of Micro-Nano Quantum Chips and Quantum Control, School of Physics, State Key Laboratory for Extreme Photonics and Instrumentation, Zhejiang University, Hangzhou 310027, China}
\affiliation{College of Optical Science and Engineering, Zhejiang University, Hangzhou 310027, China}
\author{Yi-Pu Wang}
\email{yipuwang@zju.edu.cn}
\affiliation{Zhejiang Key Laboratory of Micro-Nano Quantum Chips and Quantum Control, School of Physics, State Key Laboratory for Extreme Photonics and Instrumentation, Zhejiang University, Hangzhou 310027, China}
	
\begin{abstract}
Chiral coupling opens new avenues for controlling and exploiting light-matter interactions. We demonstrate that chiral coupling can be utilized to achieve unidirectional perfect absorption. In our experiments, chiral magnon-photon coupling is realized by coupling the magnon modes in yttrium iron garnet (YIG) spheres with spin-momentum-locked waveguide modes supported by spoof surface plasmon polaritons (SSPPs). These photon modes exhibit transverse spin, with the spin direction determined by the propagation direction. Due to the intrinsic spin properties of the magnon mode, it exclusively couples with microwaves traveling in one direction, effectively suppressing the reflection channel. Under the critical coupling condition, transmission is also eliminated, resulting in unidirectional perfect absorption. By incorporating additional YIG spheres, bidirectional and multi-frequency perfect absorption can be achieved. Our work introduces a novel platform for exploring and harnessing chiral light-matter interactions within spin-momentum locked devices, offering a paradigm for unidirectional signal processing and energy harvesting technologies.

\end{abstract}
\maketitle

Chiral light-matter interaction is a phenomenon in which the coupling between light and matter is influenced by the handedness, or chirality, of both the light and matter~\cite{Lodahl,Hentschel,Lininger-23}. It not only provides a method for sensing and control of enantiomers~\cite{Riso-23,Jiang,ZhangChi,Mayer-24}, but also supports light polarization control and detection with chiral materials~\cite{Turner,Yangying,Liwei}. From the perspective of photonic information processing, chiral coupling can induce direction-dependent photon emission, scattering, and absorption. These nonreciprocal characteristics are valuable for developing novel optical and quantum applications~\cite{Sayrin,Jalas,Scheucher,Cardano,Riccardo} across various platforms, including cold atoms~\cite{Pucher,Antoniadis}, superconducting qubits~\cite{Joshi,Owens}, quantum dots~\cite{Immo,Coles,Hurst}, plasmonic meta-atoms~\cite{Mun,Wang}, and magnonics~\cite{Wang-1,zhangxu,Weichao,Bourhill,Tao-1,Tao-2,Tao-3,zhiyuan,Jie,ziyuan,zhangchi}. Moreover, chiral light-matter interface is essential for developing complex quantum networks~\cite{Kimble,Reiserer,Pichler,Pichler,Mahmoodian}, facilitating deterministic quantum state transfer~\cite{Cirac,Vermersch}, and enabling the simulation of novel quantum many-body systems~\cite{Douglas,Anderson}.

Beyond the mentioned applications and platforms, further exploration of chiral coupling and its control techniques is both necessary and intriguing. Here, we focus on electromagnetic wave control, where perfect absorption has garnered significant interest~\cite{Chong,Baranov,yanglan,Tretyakov,Pichler2,Slobodkin,Aeschlimann,Landy,Mullers}. This is vital for applications like energy harvesting and information storage. Achieving total absorption of incident light, without reflection or transmission, is challenging due to strict parameter requirements, often requiring precise control of phase and amplitude across multiple input fields~\cite{Rao,QJ,Wan,Soleymani,Kim}. In this work, we demonstrate that chiral coupling offers a promising strategy for achieving perfect absorption, reducing the need for complex parameter control and tuning. This approach provides a simpler, potentially more robust path~\cite{Bliokh-14,Jacob-16,Bliokh-19} to perfect absorption. Notably, this approach enables \textit{unidirectional} perfect absorption, where incident fields from the opposite direction remain unabsorbed. Furthermore, this method can be easily extended to achieve bidirectional and multi-band perfect absorption, demonstrating its flexibility for practical applications.

To achieve chiral light-matter interactions, we can utilize spin-momentum-locked photon mode~\cite{Bliokh-1,Bliokh-2,Bliokh-3,Petersen,Hallett,Fortuno,Mo}, while ensuring that the interacting matter possesses spin properties or exhibits circular polarization characteristics~\cite{Lodahl}. Spin-momentum locking in photonics refers to the dependence of the transverse spin of the electromagnetic field on its propagation direction~\cite{Bliokh-14,Jacob-16}. This phenomenon is commonly observed in tightly and transversely confined optical fields, such as those in nanofibers~\cite{Sollner, Luxmoore}, whispering gallery mode resonators~\cite{Shomroni, Sayrin, Chiasera}, and surface plasmon polariton waveguides~\cite{Mechelen, Bliokh-4, Liu-1, Liu-2}. In this work, we achieve photon-magnon chiral coupling within a waveguide magnonic system, where spoof surface plasmon polaritons (SSPPs) support spin-momentum-locked waveguide modes~\cite{Pendry,Francisco,Maier}, and ferrimagnetic YIG spheres provide the magnon modes (spin collective excitation modes). The handedness of the spin precession, or the polarization of the magnon mode, is determined by the direction of the bias magnetic field. In the chiral coupling scenario, magnons couple with photons propagating in one direction, while backscattering is suppressed. We then employ the magnon-photon critical coupling condition~\cite{Cai-critical,Yu-critical,Yang-critical} to cancel transmission, resulting in unidirectional perfect absorption. This \textit{unidirectionality} arises because the photon mode traveling in the opposite direction, with an opposing transverse spin, cannot interact with the magnon, allowing unobstructed transmission.

\begin{figure*}[t!]
\includegraphics[width=0.9\textwidth]{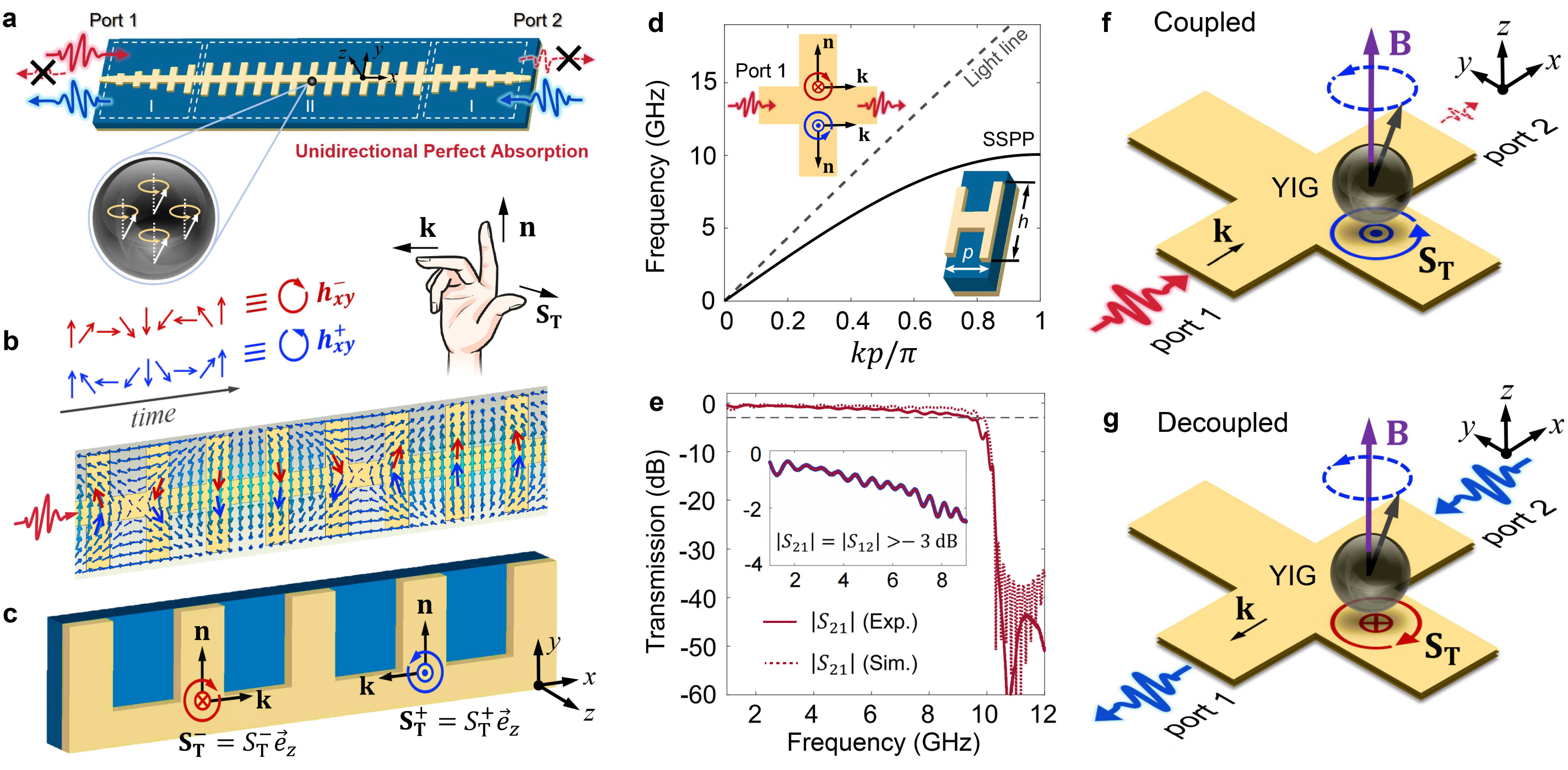}
\caption{\textbar~\textbf{Chiral coupling in spin-momentum locked waveguide magnonics.} \textbf{a} Schematic of the SSPP waveguide. Part I: the region for momentum matching transition. Part II: the region supporting the propagating SSPP mode. \textbf{b} Distribution of in-plane magnetic filed vectors, where the vectors rotate elliptically at the arrow-marked positions. \textbf{c} Schematic of the transverse spin and spin-momentum locking in the SSPP waveguide. \textbf{d} Dispersion relation of the SSPP mode (solid curve) and the light line (dashed curve). The upper inset shows the polarizations at symmetric points along the $y$-axis when the microwave is input from Port 1. The bottom insert shows one unit of the SSPP waveguide, with $p=4.1~\rm{mm}$ and height  $h=7.6~\rm{mm}$. \textbf{e} Measured (solid curve) and simulated (dashed curve) transmission spectra of the SSPP waveguide. \textbf{f, g} Schematic of the chiral coupling between the YIG sphere and the SSPP waveguide.}\label{fig1}
\end{figure*}

\begin{figure*}[t!]
	\includegraphics[width=0.9\textwidth]{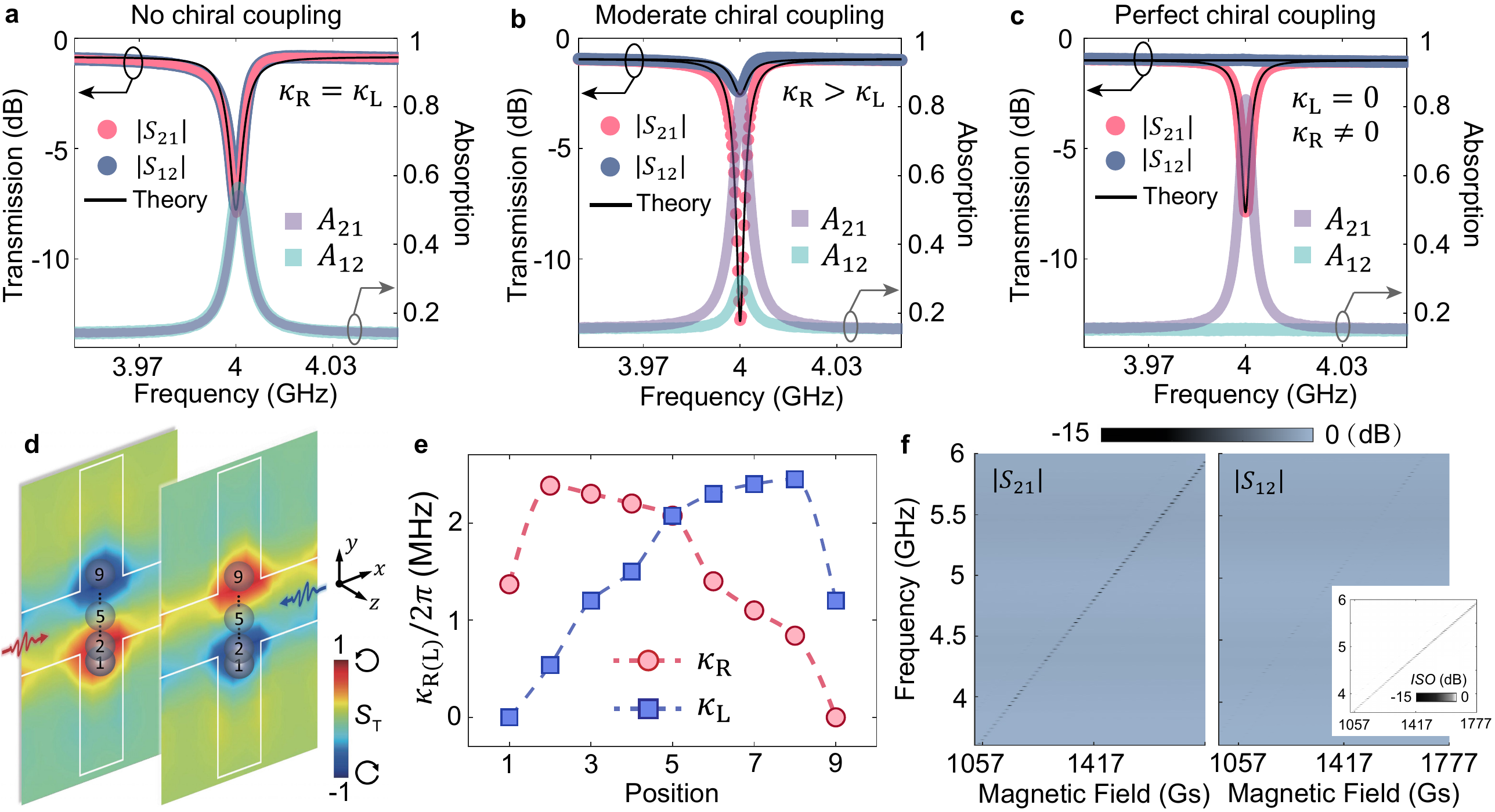}
	\caption{\textbar~\textbf{Tunable chiral couplings in experiments.} \textbf{a-c} Transmission spectra measured at three different positions of the YIG sphere (P5, P2, and P1, as marked in \textbf{d}), corresponding to non-chiral, moderate chiral and perfect chiral coupling, respectively. Rightward and leftward transmission spectra, $|S_{21}(\omega)|$ and $|S_{12}(\omega)|$, are shown as red-dot and blue-dot curves, respectively. Black solid curves represent theoretical fits using Eq.~(\ref{eq5}). The corresponding absorption spectra, $A_{21}$ and $A_{12}$, are plotted by purple- and green-squares, respectively. \textbf{d} Schematic of the YIG sphere position on the SSPP waveguide, overlaid with a color map of the calculated transverse spin angular momentum (SAM) density. \textbf{e} Extracted extrinsic damping rates of the magnon mode, $\kappa_{\rm{R}}$ and $\kappa_{\rm{L}}$, when the YIG sphere is placed at different positions. \textbf{f} Nonreciprocal transmission mapping as a function of magnetic field. The inset shows the isolation ratio (\emph{ISO}) between $|S_{21}(\omega)|$ and $|S_{12}(\omega)|$.}\label{fig2}
\end{figure*}

In our device, the operating frequency can be flexibly controlled by adjusting the external magnetic field. Bidirectional perfect absorption is achieved by introducing two YIG spheres coupled to the waveguide at different positions. Furthermore, integrating additional YIG spheres could enable broadband perfect absorption, enhancing system's versatility. Our work demonstrates that spin-momentum locked interaction is a promising approach for designing directional optical and quantum devices.\\
\\
\textbf{\large{Results}}\\
\textbf{Spin-momentum locked waveguide magnonics and theoretical model}
\\Figure~\ref{fig1}a depicts the SSPP waveguide with periodic metallic grooves (period $p=4.1~\text{mm}$, details see Supplementary Fig. S1). This structure consists of three segments, with Part II (center white dashed box in Fig.~\ref{fig1}a) featuring grooves of uniform height $h = 7.6~\text{mm}$ that support the SSPP propagating mode. The periodic grooves laterally confine the microwave fields and induce a longitudinal component of the electromagnetic field. This confinement gives rise to a propagating SSPP mode that intrinsically carries transverse spin angular momentum (SAM), $\mathbf{S_T}$, over a broad frequency range. The transverse SAM arises from the elliptical rotation of the in-plane magnetic field components, a direct consequence of the broken mirror symmetry and strong transverse confinement~\cite{Bliokh-4,Liu-1,Liu-2}. Notably, the direction of $\mathbf{S_T}$ is locked to the propagation direction of the mode, forming a spin-momentum locking relation that underpins the chiral coupling observed in our system, as illustrated in Figs.~\ref{fig1}b and 1c. Figure~\ref{fig1}b shows the distribution of microwave magnetic-field vectors simulated using the CST Microwave Studio, an electromagnetic simulation software. At the positions marked by red and blue arrows in Fig.~\ref{fig1}b, the in-plane magnetic field ($\mathbf{h_{xy}}$) vectors rotate elliptically over time, which is schematically depicted in the top panel of Fig.~\ref{fig1}b. The clockwise and counterclockwise polarizations of $\mathbf{h_{xy}^{\mp}}$ correspond to transverse spins with different orientations. For convenience, we define $\mathbf{S^-_T}=S^-_\text{T}\vec{e}_z$ and $S^-_\text{T}<0$, where $ \vec{e}_z $ is the unit vector in the $z$-direction. Then $\mathbf{S^+_T}=S^+_\text{T}\vec{e}_z$ and $S^+_\text{T}>0$. The transverse spin is tied to the wave vector $\textbf{k}$ according to the right-hand rule. Here, we take the top side of the device as an example, as shown in Fig.~\ref{fig1}c, where $\mathbf{n}$ represents the interface normal, pointing from the device center toward the boundary. It can be found that the rightward-propagating waves produce a negative transverse spin ($S^-_\text{T}<0$), while the leftward-propagating waves produce a positive transverse spin ($S^+_\text{T}>0$). Conversely, for the bottom side of the device (insert in Fig.~\ref{fig1}d), the situation is reversed due to the symmetry. To quantify the degree of circular polarization and its spatial distribution, we calculate the transverse spin angular momentum (SAM) density of the SSPP waveguide modes under both rightward and leftward microwave excitation, as shown in Fig.~\ref{fig2}d.\\
Figure~\ref{fig1}d shows the simulated dispersion of the SSPP waveguide mode (solid curve). Within the first Brillouin zone, the dispersion gradually deviates from the light line (dashed curve) and becomes nearly flat as $kp/\pi$ approaches 1. At the Brillouin zone boundary, the group velocity of the SSPP mode drops to zero at the cutoff frequency $f_{\rm{c}}/2\pi=10~\text{GHz}$. To smoothly load the signal from the device input port to the SSPP mode, we employ a gradient groove design in Part I of the device (left and right white dashed boxes in Fig.~\ref{fig1}a), with groove heights varying from $2.6~\text{mm}$ to $7.6~\text{mm}$, facilitating the momentum transition~\cite{Ma} (Supplementary Section I). As shown in Fig.~\ref{fig1}e, the measured (solid curve) and simulated (dashed curve) transmission spectra demonstrate low-loss propagation, with transmission coefficients exceeding -3 dB throughout the 1 $\sim$ 9 GHz band, indicating efficient coupling into the SSPP mode. The Hamiltonian of the SSPP mode can be expressed as
\begin{equation}\label{eq1}
{H_{\rm{w}}/\hbar ={\textstyle \sum_{k}}\omega(k)\hat{p_k}^{\dagger}p_k+ {\textstyle \sum_{k}}\omega(k)\hat{q_k}^{\dagger}q_k,}
\end{equation}
where $\omega$ is the frequency of the traveling photon mode, and
$k$ denotes the magnitude of the wave vector. $\hat{p}_k~(\hat{p}_k^\dagger)$ and $\hat{q}_k~(\hat{q}_k^\dagger)$ represent the annihilation (creation) operators for the rightward and leftward traveling photon modes, respectively, with $\left [\hat{p}_{k},\hat{p}_{k^{'}}\right] =\delta(k-k^{'})$ and $\left [\hat{q}_{k},\hat{q}_{k^{'}}\right] =\delta(k-k^{'})$.

In the experiment, we study the magnon mode in a YIG sphere (1-mm diameter), specifically the uniform spin precession mode, also known as the Kittel mode. An external magnetic field $\mathbf{B}$ is applied along the $+z$ axis, as shown in Figs.~\ref{fig1}f and 1g.  The magnon mode frequency is given by $\omega_{\rm{m}}=\gamma (B+B_{\rm{A}})$, where $\gamma$ is the gyromagnetic ratio, and $B_{\rm{A}}$ represents the anisotropy field in the sphere. The magnon-photon coupling strength $\tilde{g}(k)$ is
\begin{equation}\label{eq3}
	{\tilde{g}(\textbf{k})=-\mu_0\sqrt{\frac{\gamma M_s V_s}{2\hbar}}[\mathcal{H}_{x}(\textbf{k},\rho)-i\mathcal{H}_{y}(\textbf{k},\rho )],}
\end{equation}
where $\mu_0$ is the vacuum permeability, $M_{\rm{s}}$ and $V_{\rm{s}}$ are the saturation magnetization and volume of the YIG sphere, respectively. $\mathcal{H}_{x}(\textbf{k},\rho)$ and $\mathcal{H}_{y}(\textbf{k},\rho)$ are the $x$-direction and $y$-direction magnetic field components of the SSPP waveguide mode at Cartesian coordinate $\rho=(X, Y, Z)$. Notably, the amplitude of the in-plane ($xy$-plane) magnetic field governs the coupling strength between the magnon and photon modes, whereas the transverse spin angular momentum (SAM) density dictated by $\mathcal{H}_{x}(\textbf{k},\rho)$ and $\mathcal{H}_{y}(\textbf{k},\rho)$ determines the chirality of the coupled system, as shown in Fig.~\ref{fig2}d. This chirality arises from the local phase difference between $\mathcal{H}_x$ and $\mathcal{H}_y$, which induces elliptical polarization and gives rise to a nonzero transverse SAM. Detailed field distributions of the SSPP waveguide and their influence on magnon-photon coupling can be found in Supplementary Section V. We fix the height $Z$ of the YIG sphere relative to the device plane, as well as its $X$ position at the center of each metallic groove. In this configuration, chiral coupling is fully controlled by moving the YIG sphere along the $y$-axis. In the presence of chiral coupling, the Hamiltonian for the interaction between the spin-momentum-locked photon mode and magnon mode can be expressed as:
\begin{figure*}[t  ]
	\includegraphics[width=0.9\textwidth]{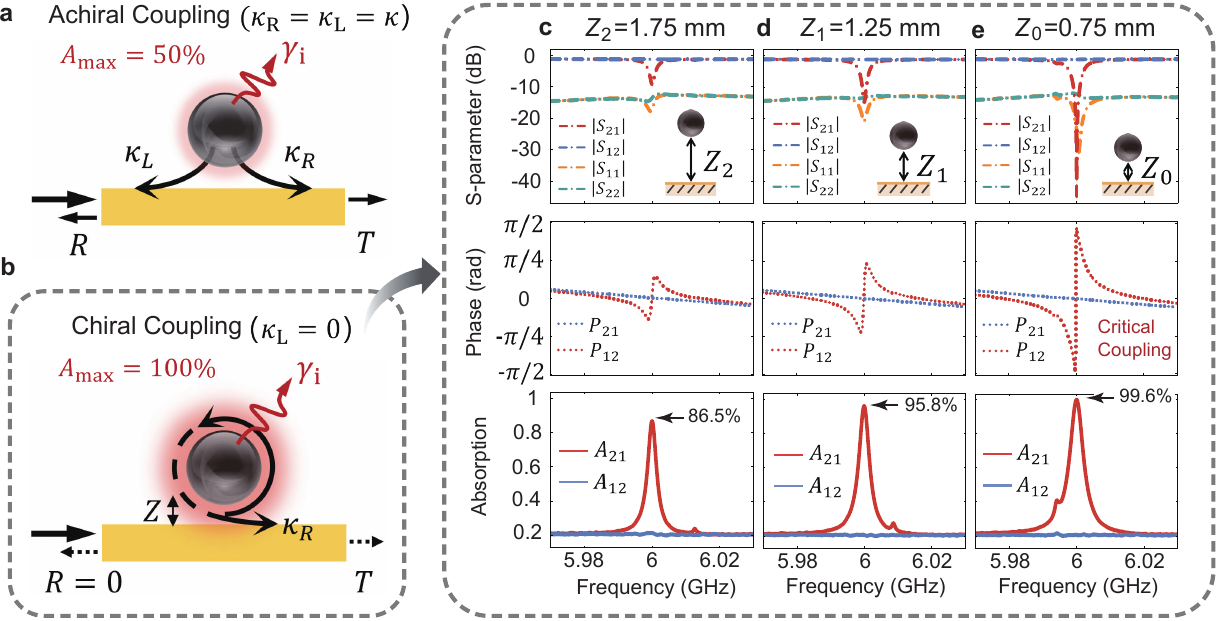}
	\caption{\textbar~\textbf{Realization of unidirectional perfect absorption.} \textbf{a, b} Schematics of achiral and chiral couplings in a two-port device. For achiral coupling, the maximum absorption rate is limited to 50\%. In contrast, under perfect chiral coupling, the absorption can reach 100\% when the critical coupling condition ($\kappa=2\gamma_{\rm{i}}$) is satisfied. \textbf{c-e} Spectral evolution as the vertical distance (\emph{Z}) between the YIG sphere and the SSPP waveguide is tuned. The first row shows the S-parameters: transmission coefficients $|S_{21(12)}(\omega)|$ plotted as red (blue) dash dot curves and reflection coefficients $|S_{11(22)}(\omega)|$ as yellow (green) dash dot curves. The second row represents the transmission phase: $P_{21}$ (red dotted curves) and $P_{12}$ (blue dotted curves) as functions of frequency. The third row shows the absorption spectra: $A_{21}$ (red curves) and $A_{12}$ (blue curves). All measurements are performed under perfect chiral coupling ($\kappa_{\rm{L}}=0$), with the YIG sphere placed at position 1 in Fig.~\ref{fig2}d. When the YIG is further brought to $Z_0=0.75~\rm{mm}$ (panel \textbf{e}), the system satisfies the critical coupling condition, and the unidirecitonal absorption reaches its maximum value.}\label{fig3}
\end{figure*}
\begin{equation}\label{eq4}
	{\hat{H}_{\rm{int}}/\hbar = {\textstyle \sum_{k}} g_{\rm{R}}(k)\hat{m}^{\dagger}\hat{p}_k+ {\textstyle \sum_{k}} g_{\rm{L}}(k)\hat{m}^{\dagger}\hat{q}_k+\rm{h.c.},}
\end{equation}
where $\hat{m}^{\dagger}$ is the magnon creation operator, $g_{\rm{R(L)}}$ is the coupling strength between the magnon mode and the rightward (leftward)-propagating photon mode, and $\rm{h.c.}$ denotes the Hermitian conjugate. The total Hamiltonian of the system is given by $\hat{H}=\hat{H}_{\rm{w}}+\hat{H}_{\rm{m}}+\hat{H}_{\rm{int}}$, where $\hat{H}_{\rm{m}}=\hbar\omega_{\rm{m}}\hat{m}^{\dagger}\hat{m}$. The device is connected to the network analyzer via port 1 and port 2, as shown in Fig.~\ref{fig1}a, for spectroscopy measurements. The transmission \(S_{21(12)}\) and reflection \(S_{11(22)}\) coefficients are derived as (Supplementary Section III)
\begin{equation}\label{eq5}
	{S_{21(12)}(\omega)=1-\frac{i\kappa_{\rm{R(L)}}}{\omega-\omega_m+i(\gamma_{\rm{i}}+\frac{\kappa_{\rm{R}}+\kappa_{\rm{L}}}{2})},}
\end{equation}
\begin{equation}\label{eq6}
	{S_{11(22)}(\omega)=-\frac{i\sqrt{\kappa_{\rm{R}}\kappa_{\rm{L}}}}{\omega-\omega_m+i(\gamma_{\rm{i}}+\frac{\kappa_{\rm{R}}+\kappa_{\rm{L}}}{2})},}
\end{equation}
where $\gamma_{\rm{i}}$ denotes the intrinsic damping rate of the magnon mode, characterizing its internal losses, and $\kappa_{\rm{R(L)}}=2\pi g^2_{\rm{R(L)}}$ represents the extrinsic damping rate associated with the coupling to the rightward (leftward) propagating photon mode. These damping parameters directly determine the spectral linewidths and resonance depths in the transmission and reflection coefficients, as described by Eqs.~(\ref{eq5}) and (\ref{eq6}).
\begin{figure*}[t!]
	\includegraphics[width=0.9\textwidth]{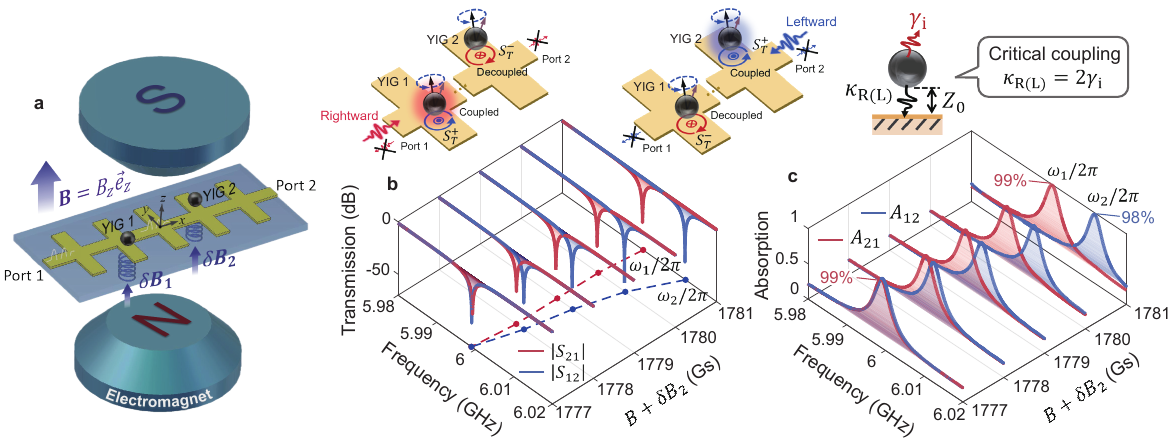}
	\caption{\textbar~\textbf{Multi-color and bidirectional perfect absorptions.} \textbf{a} Schematic diagram of the experimental setup. Two YIG spheres are integrated at different positions along the SSPP waveguide and subjected to a common vertical bias magnetic field ($\rm{\textbf{B}}$-field). The magnon mode frequencies of the two YIG spheres are independently tuned via local magnetic field adjustments ($\rm{\textbf{B}}$+$\delta\rm{\textbf{B}}_\textbf{1(2)}$) using two separate coils. \textbf{b, c} Evolution of the measured transmission and absorption spectra when the magnon mode frequency of YIG 2 is tuned. Red and blue curves correspond to rightward and leftward propagating microwaves, respectively. The top inset schematically illustrates the transverse spin directions of the SSPP waveguide modes at the locations of the two YIG spheres for each propagation direction. All spectra are measured under conditions of perfect chiral coupling and critical coupling for both YIG spheres.}\label{fig4}
\end{figure*}\\
\\
\textbf{Observation of chirality induced nonreciprocity}\\
To clearly observe the chiral coupling effect in experiment, we use a 1-mm-diameter YIG sphere that precisely overlaps with the high-density region of transverse spin, as indicated by the red and blue regions in Fig.~\ref{fig2}d. When the sphere is positioned off-center along the $y$-axis, the local transverse SAM density $S_{\rm{T}} \ne 0$ induces a directional magnon-photon interaction. For instance, placing the YIG sphere at position P1 (marked in Fig.~\ref{fig2}d) and applying an external magnetic field of $B = 1167~\rm{Gs}$ along the $+z$ direction leads to chiral coupling. As shown in Figs.~\ref{fig1}f and 1g, the counterclockwise precession of magnetization allows the magnon mode to couple exclusively with rightward-propagating photons (Fig.\ref{fig1}f), while remaining decoupled from the leftward-propagating photons due to their opposite circular polarization (Fig.~\ref{fig1}g). This chiral coupling results in the nonreciprocal transmission of the incident field. The measured transmission coefficients $|S_{21}(\omega)|$ and $|S_{12}(\omega)|$ at three typical positions (P5, P2, P1 marked in Fig.~\ref{fig2}d) are plotted in Figs.~\ref{fig2}a-2c. At P5 (the center of the $y$-axis, $Y=0$), the transverse SAM density vanishes  ($S_{\rm{T}}=0$), leading to equal transmission amplitudes $|S_{21}(\omega)| = |S_{12}(\omega)|$, as shown in Fig.~\ref{fig2}a. This indicates achiral magnon-photon coupling with $\kappa_{\rm{R}} = \kappa_{\rm{L}}$. In contrast, when the YIG sphere is placed at off-center positions such as P2 or P1, where $S_{\rm{T}} \ne 0$, chiral coupling emerges, resulting in nonreciprocal transmission: $|S_{21}(\omega)| \ne |S_{12}(\omega)|$. At the intermediate position P2, we find $\kappa_{\rm{R}} > \kappa_{\rm{L}} \neq 0$, where \textit{moderate} chiral coupling is observed, as depicted in Fig.~\ref{fig2}b. We quantify the chirality of the coupling by defining a dimensionless parameter $\mathcal{C} = (\kappa_{\rm{R}} - \kappa_{\rm{L}})/(\kappa_{\rm{R}} + \kappa_{\rm{L}})$, which characterizes the degree of asymmetry between the magnon's coupling to rightward and leftward propagating modes. By moving the YIG sphere along the $y$-axis, the chirality paraneter $\mathcal{C}$ can be continuously tuned. At position P1, $\kappa_{\rm{L}}$ is suppressed to zero, resulting in $\mathcal{C} = 1$. In this case, the field incident from the right (i.e., the leftward-propagating wave) is entirely decoupled from the magnon mode, and the transmission remains unity (0 dB), as indicated by the blue-dot curve in Fig.~\ref{fig2}c. In contrast, when the incident field is reversed, magnon-photon coupling occurs and ferromagnetic resonance is observed, as shown by the red-dot curve in Fig.~\ref{fig2}c. This unidirectional coupling behavior characterized by $|\mathcal{C}| = 1$, defines the regime of \textit{perfect} chiral coupling, where the interaction is fully suppressed in one direction while preserved in the other. The black solid curves represent the theoretical results based on Eq.~(\ref{eq5}). From the fitting, we obtain $\kappa_{\rm{R}}/2\pi=1.37~\rm{MHz}$, and $\kappa_{\rm{L}}/2\pi=0$. Figure \ref{fig2}e presents the extracted extrinsic damping rates of the magnon mode, fitted at various positions of the YIG sphere along the $y$-axis, illustrating how the coupling asymmetry evolves with position. Notably, the transverse spin of the photon mode exhibits opposite signs on either side of the device with respect to the center at $ Y = 0 $, as shown in Fig.~\ref{fig2}d. As a result, for positions with $Y < 0$ (from P1 to P5), we observe $\kappa_{\rm{L}} < \kappa_{\rm{R}}$ and hence $\mathcal{C}>0$, while for $Y > 0$ (from P5 to P9), the relation reverses with $\kappa_{\rm{L}} > \kappa_{\rm{R}}$ and $\mathcal{C}<0$. The chiral coupling-induced nonreciprocal transmission can also be achieved over a wide frequency range by simply sweeping the bias magnetic field, as illustrated by the transmission mappings of $|S_{21}(\omega)|$ and $|S_{12}(\omega)|$ in Fig.~\ref{fig2}f. The inset shows the isolation (in decibels) between the rightward and leftward transmission, defined as $\emph{ISO} = 20 \log_{10}(|S_{21}|/|S_{12}|)$.\\
\\
\textbf{Realization of unidirectional perfect absorption}\\
Next, we utilize chiral coupling, combined with the magnon-photon critical coupling condition, to achieve unidirectional perfect absorption. Based on Eqs.~(\ref{eq5}) and (\ref{eq6}), we consider the absorption rate $A_{21(12)} =1-|S_{21(12)}|^2-|S_{11(22)}|^2$ under both achiral and chiral coupling conditions. For achiral coupling (Fig.~\ref{fig3}a), where $\kappa_{\rm{R}}=\kappa_{\rm{L}}=\kappa$, the maximum absorption rate is achieved at the magnon mode resonance frequency ($\omega=\omega_{\rm{m}}$) with $\kappa=\gamma_{\rm{i}}$. In this scenario, the maximum absorption is $A_{\rm{max}}=1-2\gamma_{\rm{i}}\kappa/(\gamma_{\rm{i}}+\kappa)^2=50\%$, which represents the theoretical \textit{upper limit} for a conventional two-port device (Supplementary Section IV). However, in the chiral coupling system, as shown in Fig.~\ref{fig3}b, the backscattering is blocked ($\kappa_{\rm{L}}=0$) and the absorption rate $A_{21}=1-|S_{21}|^2=1-[(2\gamma_{\rm{i}}-\kappa_{\rm{R}})/(2\gamma_{\rm{i}}+\kappa_{\rm{R}})]^2$ can reach 100\% under the critical coupling condition ($\kappa_{\rm{R}}=2\gamma_{\rm{i}}$)~\cite{Cai-critical,Yu-critical}. Chirality enhances absorption in one direction while suppressing it in the opposite. As shown in the measured absorption spectra in Figs.~\ref{fig2}a-2c, when $\kappa_{\rm{R}} > \kappa_{\rm{L}}$ ($\mathcal{C} > 0$), the rightward absorption $A_{21}$ (Figs.~\ref{fig2}b and~\ref{fig2}c) exceeds the reciprocal case shown in Fig.~\ref{fig2}a, whereas the leftward absorption $A_{12}$ is correspondingly reduced. When the system is tuned to the regime of \textit{perfect} chiral coupling ($\mathcal{C} = 1$ in Fig.~\ref{fig2}c), the leftward absorption is zero.

In the following experiment, we fix the YIG sphere at position P1 to ensure \textit{perfect} chiral coupling ($\kappa_{\rm{L}} = 0$ and $\mathcal{C} = 1$), and finely adjust its vertical distance $Z$ from the device plane to tune $\kappa_{\rm{R}}$ accordingly, aiming to optimize the rightward absorption $A_{21}$. The measurements are performed near $6~\mathrm{GHz}$ as a typical example within the tunable range of the magnon mode. This choice is not essential, as the magnon resonance frequency in the YIG sphere can be continuously adjusted via the external magnetic field, allowing the demonstrated unidirectional absorption to be readily extended to other frequency ranges.

Figures~\ref{fig3}c-3e show the spectral evolution as the YIG sphere is gradually brought closer to the device plane. The first row presents the measured $S$-parameters. The reflection coefficients $|S_{11}(\omega)|$ (yellow dash dot curves) and $|S_{22}(\omega)|$ (green dash dot curves) remain around $-15~\mathrm{dB}$ across the frequency range. The perfect chiral magnon-photon coupling further suppresses the reflected signal at the resonance frequency.

The second and third rows display the transmission phase and absorption for rightward and leftward propagating waves, respectively. As the YIG sphere moves from $Z_2$ to $Z_0$, $\kappa_{\rm{R}}$ increases, leading to a broadening of the $|S_{21}(\omega)|$ (red dash dot curves) and $A_{21}$ (red curves) spectra. Correspondingly, $|S_{21}(\omega)|$ decreases at the magnon resonance frequency $\omega/2\pi = 6~\mathrm{GHz}$. When the critical coupling condition is satisfied, i.e., $\kappa_{\rm{R}}/2\pi = 2\gamma_{\rm{i}}/2\pi = 2.4~\mathrm{MHz}$, a pronounced dip with a $\pi$ phase shift appears in $|S_{21}(\omega)|$, as shown in Fig.~\ref{fig3}e (see Supplementary Section IV for details). At this point, the rightward-propagating microwaves are nearly perfectly absorbed by the magnon mode, with $A_{21} = 99.6\%$ at $\omega_{\rm{m}}$. In contrast, the reverse transmission amplitude $|S_{12}(\omega)|$ (blue dash dot curves) and the corresponding phase $P_{12}$ (blue dotted curves) remain unchanged, indicating that the magnon mode remains fully decoupled from the leftward-propagating wave. In addition, several minor absorption peaks can be observed near the Kittel mode frequency in the absorption spectra, which originate from higher-order magnetostatic modes of the YIG sphere. These modes exhibit relatively weak coupling to the waveguide mode and are not optimized for signal absorption, resulting in low absorption amplitudes. Therefore, they are neglected in the context of perfect absorption work. However, considering that these higher-order modes carry additional degrees of freedom such as orbital angular momentum and complex spin textures~\cite{Yasu2019}, they hold potential value for future exploration in the development of chiral microwave devices.\\
\\
\textbf{System tunability and extendibility}\\
Finally, we demonstrate the tunability and extendibility of the system, which includes adjusting the frequency at which perfect absorption occurs and achieving bidirectional perfect absorption. These objectives can be conveniently realized by tuning the bias magnetic field and integrating additional YIG spheres. As shown in Fig.~\ref{fig4}a and in the top inset of Fig.~\ref{fig4}b, two YIG spheres are placed at symmetric points along the $y$-axis. For rightward propagating waves, microwave mode at the lower (upper) point exhibits positive (negative) $S_\text{T}$. For leftward propagating waves, the situation reverses. In the experiment, a global bias field ($\mathbf{B}$-field) is applied perpendicularly to the device plane, which saturates the magnetizations of the YIG spheres. To independently fine-tune the resonance frequency of the magnon mode in each YIG sphere, we place a small electromagnet coil beneath each sphere as shown in Fig.~\ref{fig4}a. The magnetic field direction of the small electromagnet is parallel to the $\mathbf{B}$-field. When the $\mathbf{B}$-field is applied along the $+z$ direction ($B_{z}>0$), magnon mode 1 of frequency $\omega_1$ in YIG sphere 1 couples to the rightward propagating waves but not to the leftward propagating waves. This yields $|S_{21}(\omega_1)|<1$ and $|S_{12}(\omega_1)|=1$ at the resonance frequency $\omega=\omega_1$. For magnon mode 2 in YIG sphere 2, the situation is reversed, such that at its resonance frequency $\omega = \omega_2$, $|S_{21}(\omega_2)| = 1$ and $|S_{12}(\omega_2)| < 1$. The measured spectra of $|S_{21}(\omega)|$ and $|S_{12}(\omega)|$ are shown in Fig.~\ref{fig4}b as the red and blue curves, respectively. Under critical coupling conditions (top inset of Fig.~\ref{fig4}c), unidirectional perfect absorption occurs for both left- and right-incident fields, at two \textit{distinct} frequencies, $\omega_1$ and $\omega_2$. This could refer to multi-color perfect absorption. By tuning the small coil magnet to have $\omega_1 = \omega_2$, we achieve bidirectional perfect absorption, as shown in Fig.~\ref{fig4}c. Regardless of which port the incident field enters from, it can be approximately 99\% absorbed around 6 GHz. To achieve perfect absorption over a broader frequency range, additional YIG spheres can be integrated and tuned to cover a continuous frequency band. Additionally, the absorption direction of each YIG sphere can be controlled by the orientation of the external magnetic field. For example, reversing the $\mathbf{B}$-field to the $-z$ direction ($B_{z} < 0$) results in a reversal of the directional response, effectively swapping the behaviors shown in Figs.~\ref{fig4}b and \ref{fig4}c. Further details are provided in Supplementary Section II.\\
\\
\textbf{\large{Discussions}}
\\We have achieved chiral coupling between microwaves and magnon modes using a coupled SSPP-YIG system. Building upon the enhanced magnon-photon coupling established in SSPP-magnon hybrid systems~\cite{Zhangxufeng}, our work further explores and utilizes the microwave spin degree of freedom inherent in the SSPP waveguide. In this hybrid system, the magnons couple solely with microwaves propagating in one direction, while the reflection channel is completely closed. Furthermore, by leveraging the critical coupling condition between the magnon mode and the photon mode, the transmitted field is entirely suppressed through the Fabry-P$\rm{\acute{e}}$rot interference, resulting in unidirectional perfect absorption.

Unlike the chiral bound states formed through three-mode hybridization in cavity magnonic systems~\cite{Han}, the chiral photonic state demonstrated in this work originates solely from the surface-propagating photon mode. As a result, it offers a simpler and more flexible approach to realizing chiral magnon-photon coupling over a broad microwave frequency range, without the need for precise control of multiple coupled modes. Owing to the tunability of the magnon mode, the frequency of perfect absorption can be flexibly adjusted by the bias magnetic field. By integrating additional YIG spheres, we demonstrate bidirectional perfect absorption, showcasing system's extendability. Our work demonstrates the potential of chiral light-matter interactions for developing directional signal processing and energy harvesting devices. It may also provides new ideas to design advanced spin-momentum locking devices.\\
%We develop new methods and paradigms for achieving perfect absorption, presenting a highly scalable absorber device that offers new avenues for energy harvesting.Future research could focus on leveraging chiral magnon-photon coupling to enable directional information processing and to develop advanced quantum information schemes.
\\
\textbf{\large{Methods}}\\
\textbf{Device design}\\
As illustrated in Supplementary Section I, we fabricated the spoof surface plasmon polariton (SSPP) waveguide on a $10\times2~\rm{cm^2}$ RO4003C substrate. The RO4003C substrate has a thickness of 0.813 mm and a dielectric constant of 3.38$\pm$0.05, the copper layers are 0.035 mm thick. Below the cutoff frequency $10~\rm{GHz}$, microwaves propagate well in the SSPP waveguide, and the insertion loss is less than $3~\rm{dB}$ in the range of $1\sim9~\text{GHz}$. In the SSPP waveguide, the nonzero transverse spin ($S_{\rm{T}}$) is primarily concentrated within an area of $1.6 \times 1.6~\mathrm{mm}^2$. To ensure good spatial mode overlap and polarization matching between the magnon and the chiral photon mode, a YIG sphere with a diameter of $1~\mathrm{mm}$ is placed to cover the high-$S_{\rm{T}}$ region. The bias magnetic field applied to the YIG sphere is aligned with the direction of $S_{\rm{T}}$.
In the experiment to demonstrate system tunability and extendibility, two $1~\rm{mm}$-diameter yttrium ion garnet (YIG) spheres are glued to the end of a displacement cantilever and adjusted accurately close to the planar device through a three-dimensional displacement motor. Two small electromagnetic coils are positioned underneath the YIG spheres, locally adjusting the magnetic field at each YIG sphere's position.\\
\\
\textbf{Measurement setup}\\
All measurements were conducted at room temperature. The end of the SSPP device are connected to a vector network analyzer (Keysight E5080B) to obtain the scattering parameters of the waveguide magnonic system. The excitation power is maintained at $-5~\mathrm{dBm}$ throughout all measurements to ensure the magnon mode operates within its linear frequency response regime.\\
\\
\textbf{\large{Data availability}}\\
The main data supporting the findings of this study are available within the article and its Supplementary Figures. The source data underlying Figs.~\ref{fig1}-\ref{fig4} of the main text are provided as a Source Data file. Additional details on datasets that support the findings of this study will be made available by the corresponding author upon reasonable request. Source data are provided with this paper.\\
\\
\textbf{\large{Code availability}}\\
The simulation and computational codes of this study are available from the corresponding author upon reasonable request.\\
\\
\textbf{\large{References}}\\

\space
\textbf{\large{Acknoledgments}}\\
This work is supported by the National Key Research and Development Program of China ({NO.~2023YFA1406703} and No.~2022YFA1405200), National Natural Science Foundation of China (No.~$92265202$, No.~$11934010$, and No.~$12174329$).\\
\\
\textbf{\large{Author contributions}}\\
Y.P.W. and J.Q. conceived the idea and initiated the research project. J.Q. designed the samples and performed the experiments with input from Y.P.W. J.Q. and Y.P.W. carried out the data analysis. J.Q, Y.P.W, J.Q.Y and C.M.H. drafted the manuscript. Q.H., Z.Y.W., W.X.W. and Y.H.Y. were involved in discussion of results and the final manuscript editing. Y.P.W., C.M.H., and J.Q.Y. supervised the project.\\
\\
\textbf{\large{Competing interests}}\\
The authors declare no competing interests.\\

\end{document}